\documentclass[journal]{IEEEtran}
\usepackage{amsmath,amsfonts}
\usepackage{algorithmic}
\usepackage{algorithm}
\usepackage{array}
\usepackage[caption=false,font=footnotesize,labelfont=rm,textfont=rm]{subfig}
\usepackage{textcomp}
\usepackage{stfloats}
\usepackage{url}
\usepackage{verbatim}
\usepackage{graphicx}
\usepackage{cite}
\usepackage[hidelinks]{hyperref}
\usepackage{hyperref}
\usepackage{booktabs}
\usepackage{url}
\usepackage{soul, color, xcolor}

\begin{document}

\title{Grating haptic perception through touchscreen: Sighted vs. Visually Impaired}

\author{Yichen Gao, Menghan Hu,~\IEEEmembership{Senior Member,~IEEE,} Gang Luo}
\maketitle

\begin{abstract}
In this study, we investigate the feasibility of enabling visually impaired (VI) individuals to comprehend graphs through haptic feedback on smartphone touchscreens using an experimental setup that closely approximates everyday mobile device usage in VI and sighted (S) individuals. Six VI and 10 S participants participated in two experiments designed to compare their ability to interpret grating images with a finger swiping across a smartphone touchscreen without vision. Vibration is triggered when a swipe gesture crosses black stripes. The participants were tasked with: (1) determining whether a stripe pattern was presented on the touchscreen and (2) comparing repeated stripe patterns of two different widths to identify which had a wider spacing. Results demonstrated that the VI group exhibited superior vibrotactile discrimination compared with the S group, as evidenced by their significantly better performance in Experiment 1 (accuracy of 99.15\% vs. 84.5\%). Experiment 2 revealed that the peak performance of the VI participants was approximately 0.270 cycles/mm (83.3\% accuracy), a frequency similar to Braille dot spacing, whereas the S group peaked around 0.963 cycles/mm (70\% accuracy). These findings suggest that coded tactile stimulation could be potentially used to present an interpretable graph for the VI. Such an approach could offer value to research in human–computer interaction and sensory adaptation.
\end{abstract}


\section{Introduction}
\IEEEPARstart{A}{s} of 2020, an estimated 1.1 billion people worldwide are affected by vision loss, including 43 million who are completely blind. In the absence or reduction of visual perception, individuals with visual impairments must rely on other sensory modalities, supported by assistive technologies, to access visual information. Touch serves as a critical sensory channel in daily life, work, and education. The hands—particularly the fingertips—are represented by disproportionately large areas in the somatosensory cortex, reflecting their importance in tactile processing. For individuals who have lost their vision, touch becomes an essential means of perceiving and interacting with their environment. Therefore, employing tactile perception as a substitute for certain aspects of visual perception represents a reasonable and effective strategy for enabling visually impaired (VI) individuals to access visual information.

Recently, numerous haptic-based innovations have emerged to enhance mobile accessibility for individuals with visual impairments~\cite{EHR11, EHR12, EHR13}. A foundational development in this domain was the creation of the Optacon in 1971, a device that allows VI users to read printed text by converting visual characters into tactile stimuli via a matrix of vibrating pins~\cite{EHR6, EHR7}. Building on this foundational work, research on refreshable tactile displays explores the emerging potential as an innovative solution for improving graphical accessibility for individuals who are blind or have low vision~\cite{rfrs,rfrs2}. The dynamic hardware devices, typically comprising pin-based actuated surfaces, enable the tactile rendering of spatial and graphical content, including Braille and complex data visualizations, and support real-time updates, user interaction, and the possibility of user-generated content. Tactile perception is also being explored through other haptic modalities, such as vibrotactile feedback. In 2021, a research team from Norway developed and evaluated a prototype vibro–audio map designed to convey complex graphical information through nonvisual means~\cite{map}. This multisensory interface uses haptic feedback on vibratory touchscreens combined with auditory cues. This research provides compelling evidence for the effectiveness of multimodal systems that integrate haptic and auditory feedback for nonvisual spatial information transmission.

However, the reliance on dedicated haptic hardware introduces critical limitations. Specifically, the dependence on specialized equipment and the frequent need for controlled usage environments restrict the technology’s immediate scalability and practical deployment.

The limitations of specialized hardware underscore the importance of leveraging ubiquitous platforms, such as smartphones. Smartphones have evolved into indispensable visual aids in this era of digital connectivity~\cite{EHR2, EHR3}. A notable rise in touchscreen-enabled smart device adoption among individuals with visual impairments has been observed, with usage rates escalating from 12\% (2009) to 82\% (2014)~\cite{lou1}. Nevertheless, a gap persists between high usage and high usability. Recent research on Saudi Arabian banking applications highlighted this gap, indicating that none of the evaluated applications conformed to minimum accessibility standards~\cite{Alayed2025}. This persistent failure in basic accessibility drives the need for more advanced assistive technologies.

In 2010, researchers from the University of Washington proposed V-braille, a novel application that leverages mainstream smartphone touchscreens and vibration feedback to emulate traditional six-dot Braille cells~\cite{EHR14}. By dividing the screen into six responsive zones, correlating with Braille dots, selective vibration intensity aids in distinguishing raised dots, enabling tactile reading without necessitating multitouch capabilities. Although V-Braille meets the needs of VI people for “nonspecialized” devices, it can only present a Braille letter once on the screen to ensure a correct recognition rate.~\cite{EHR14}.

Besides the aforementioned studies, as evident from the Nokia Braille Reader prototype and tactile Braille reading with a stylus from Rantala et al.~\cite{EHR9, EHR10}, most commercially available or near-commercial haptic-based innovations have mainly focused on the tactile recognition of alphanumeric characters, including standard print and Braille. However, beyond textual content, graphical information constitutes a critical component of visual perception. Accordingly, it is meaningful to develop assistive technologies that facilitate access to graphical content for individuals with visual impairments.

A commonly used assistive technology for conveying visual content is alternative text (ALT)~\cite{alt}, which involves third-party textual descriptions of images, which are typically delivered to VI users via screen readers or text-to-speech systems. However, ALT is not a “direct” perceptual modality; it relies on an intermediary’s interpretation of the visual source, which inherently introduces the risk of information distortion or loss.

Although auditory substitution is a widely used approach for vision assistance, it is not suitable for people with hearing loss~\cite{EHR5}. For example, Usher syndrome, a genetic disorder, can cause combined hearing and vision loss~\cite{Usher}.

Touch is a form of multimodal perception, as several different receptors are integrated into a smaller piece of the receptive field. Merkel nerve endings are mechanoreceptors, a class of sensory receptors, found in the basal epidermis and hair follicles. They provide information on mechanical pressure, position, and deep static touch features, such as shapes and edges~\cite{In1, In2}. The Pacinian corpuscle is a type of mechanoreceptor found in hairy and hairless skin, viscera, and joints; it is attached to the bone’s periosteum. It is primarily responsible for vibration sensitivity~\cite{In3}. A vibratory stimulus can be conceptualized as a static touch stimulus that varies regularly in time. This is because Merkel receptors are responsive to pressure, whereas Pacini receptors are responsive to pressure application and subsequent release. This multimodal form is crucial in humans’ perception of object texture. Merkel receptors collect spatial cues provided by relatively large surface elements, such as bumps and grooves, which can be perceived through tactile interaction. Such spatial information can be detected when the skin is in motion across the surface elements and when it is in contact with them. Meanwhile, Pacinian receptors receive temporal cues when the skin moves across a textured surface, such as fine sandpaper. This type of cue provides information in the form of vibrations resulting from movement over the surface. Temporal cues are crucial for perceiving fine textures, which cannot be detected unless the fingers are in motion across the surface. This dual-mode approach markedly enhances touch perception accuracy and information acquisition scope.

\begin{figure}[!t]
\centering
\subfloat[]{\includegraphics[width=0.5\columnwidth]{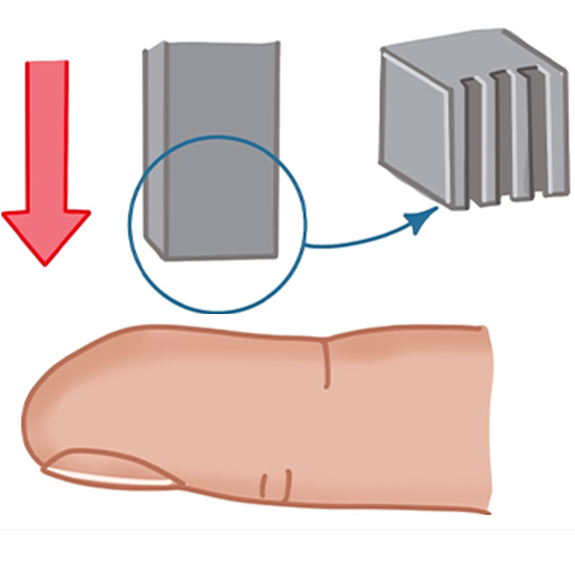}%
\label{finger1}}
\hfil
\subfloat[]{\includegraphics[width=0.5\columnwidth]{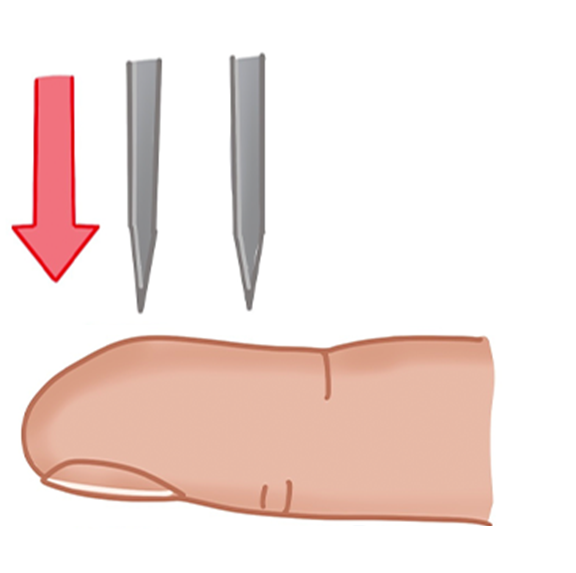}%
\label{finger2}}
\caption{Method for determine tactile acuity: two-point threshold and grating acuity. Redrawn from $Sensation$ $and$ $Perception$, page 362~\cite{book}. (a) Determine whether it is one or two points. (b) Determine whether the grating is oriented horizontally or vertically.}
\label{finger}
\end{figure}

In response to the aforementioned characteristics of haptics, there are two principal methods for detecting tactile acuity. The most established method for measuring tactile acuity is the two-point threshold, which represents the minimum separation between two points on the skin that are perceived as two distinct points when stimulated~\cite{In4}. The two-point threshold is determined by gently touching the skin with two points, such as the points of a drawing compass, and inquiring from the participants whether they perceive one or two points (Fig.~\ref{finger1}). The other method is grating acuity measurement, which is accomplished by applying a grooved stimulus to the skin and instructing the participants to identify the grating orientation, whether it is vertical or horizontal (Fig.~\ref{finger2})~\cite{In5}. Acuity is quantified by determining the narrowest spacing for which orientation can be accurately discerned. In addition, acuity can be gauged by pushing raised patterns, such as letters, onto the skin and determining the smallest pattern or letter that can be identified. Further, both tasks are conducted under passive touch conditions.

The extant research indicates that the prevailing methods for assessing tactile acuity predominantly focus on Merkel receptors to evaluate the sensitivity of participants to the surfaces of static objects. Concurrently, research has demonstrated that the prevailing tactile acuity testing methods are frequently unduly reliant on tangible materials, and less comparative testing of visually related variables has been performed in the experiment~\cite{RVHap, atkins}. Notably, tangible materials are primarily perceived through mechanoreceptors activated by pressure-induced skin displacement~\cite{lou2}, whereas tactile interactions on a touchscreen are not pressure-driven but mainly involve the stimulation of vibration-sensitive Pacinian corpuscles, which exhibit maximal neural innervation in the 200–300-Hz range~\cite{lou3, lou4}. This further distinguishes our approach from existing methods.

In accordance with the aforementioned requirements, this pilot study explores the feasibility and preliminary efficacy of a smartphone-operated visual assistance system to transmit graphical information through tactile vibration feedback. The primary goal of this initial investigation is to assess the potential of such a system to be portable, widely accessible, and capable of effectively conveying graphical information.

To preliminarily evaluate the proposed design, we adopt a dual-experiment framework to explore: (1) whether blind subjects can effectively acquire information through the tactile screen, (2) whether blind subjects might exhibit superior tactile discrimination capabilities over sighted (S) counterparts in haptic perception tasks, and (3) the unique characteristics and tactile strategies exhibited by blind individuals when interpreting vibrotactile information. Rooted in neuroplasticity~\cite{rv1, rv2, rv3}, we hypothesize that vibrotactile image perception is feasible for VI individuals, as their daily use of touch enhances their tactile acuity and provides clear advantages in vibration recognition. Concurrently, the experimental design incorporates comparative tactile feature extraction to characterize the unique haptic processing signatures of the VI population. These preliminary empirical findings aim to provide early insights into optimizing haptic interfaces tailored to the sensory strengths of VI users. The implementation of such vibration-encoded assistive technology demonstrates the potential for advancing accessible graphical interpretation systems, and this pilot study lays the groundwork for future studies aimed at enhancing the accessibility and quality of life for VI individuals.

\section{Methodology}

In light of previous research, we introduce a two-part experiment. Drawing on standard grating sensitivity measurements, a smartphone-based vibrotactile discrimination test was developed. Unlike traditional studies that focus on passive touch, we emphasize active touch, as conveying image information effectively requires users to actively explore the stimuli, with active touch outperforming passive touch in information recognition and acquisition. In addition, custom-designed images were used to replace the physical protrusions typically employed in conventional tactile experiments.~\cite{rv4}

The core design principles of this study are as follows: the experimental system constructs a haptic‐feedback mapping by analyzing the pixel distribution of binarized images. Specifically, as participants explore an image via the touchscreen interface, the system continuously monitors the pixel value at the touch coordinates. Contact with a black pixel (RGB 0,0,0) triggers an initial vibrotactile pulse. The subsequent feedback depends on finger movement: sliding across black pixels elicits continuous vibration, whereas moving onto white pixels (RGB 255,255,255) produces no feedback. If the touch remains stationary, the vibration ceases after the initial pulse. Ultimately, the participants used these haptic cues to infer the image content. This dynamic, content‐driven haptic encoding strategy conveys the spatial structure of graphical information through tactile modality. During the experiment, participants—deprived of visual assistance—actively explored an image to obtain vibratory cues and then formed internal representations to interpret the image content based on the tactile experiences. Researchers recorded key metrics, including recognition accuracy, task completion time, and exploration path trajectories. Through quantitative analysis of multidimensional behavioral data, we aim to elucidate how haptic encoding strategies affect the efficiency of graphical information transmission and provide empirical evidence for optimizing nonvisual human–computer interaction systems.

\begin{figure}[h]
\centering
\subfloat[]{\includegraphics[width=0.33\columnwidth]{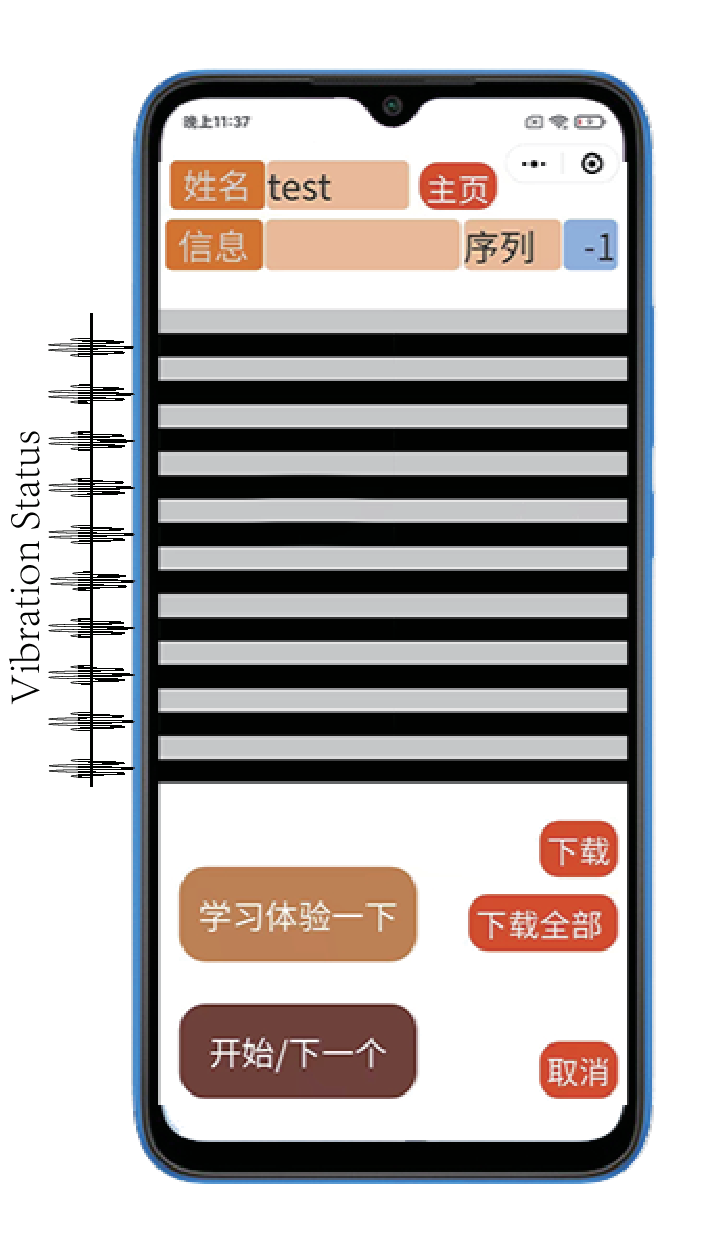}%
\label{exp1_1}}
\hfil
\subfloat[]{\includegraphics[width=0.33\columnwidth]{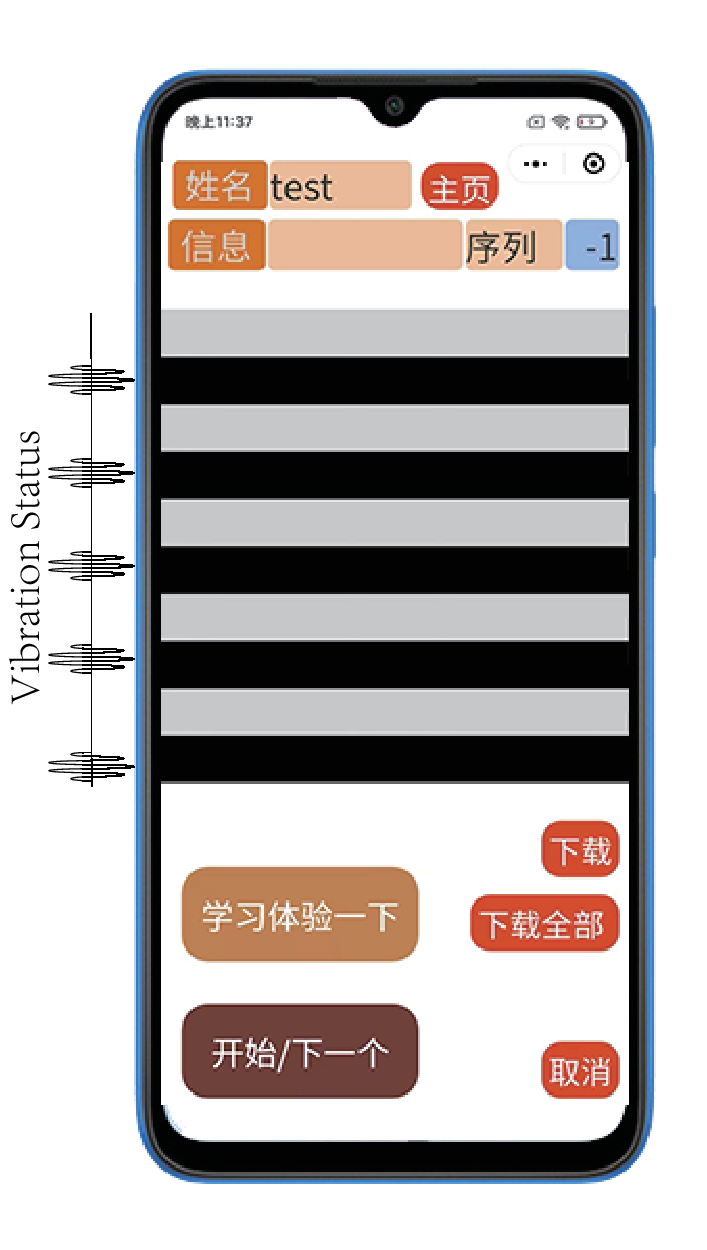}%
\label{exp1_2}}
\hfil
\subfloat[]{\includegraphics[width=0.32\columnwidth]{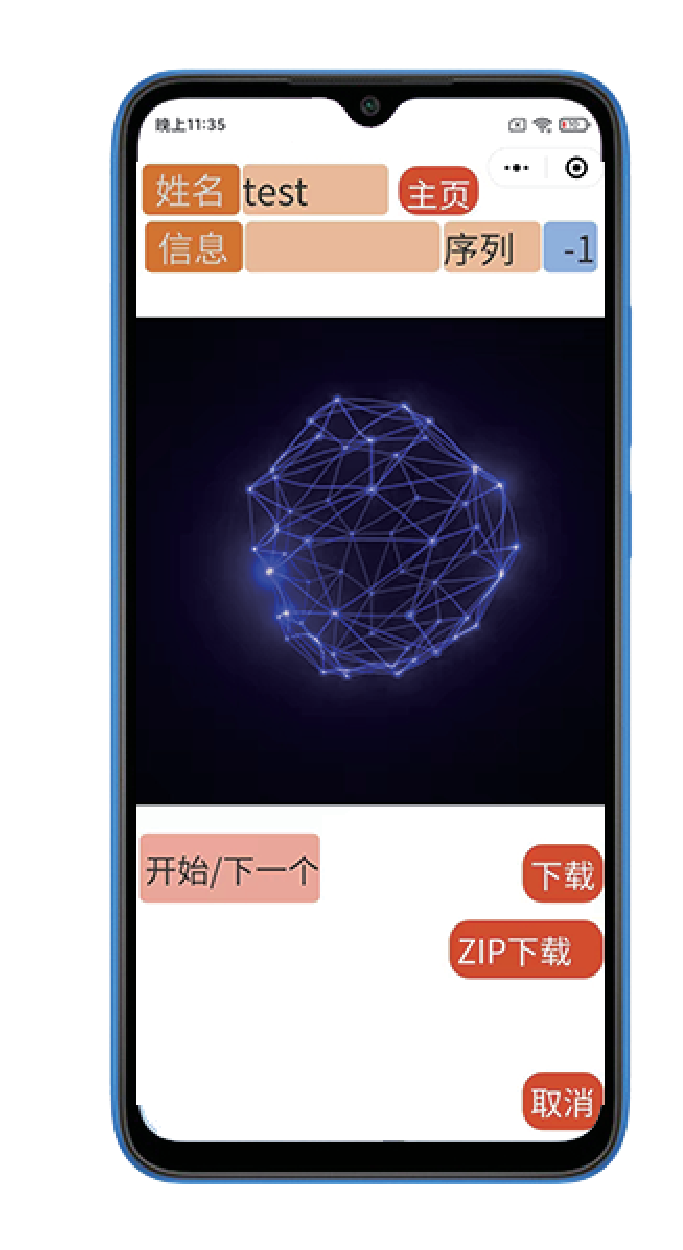}%
\label{exp1_3}}
\caption{Smartphone interface for Experiment 1. (a) and (b) illustrate two examples of the grating stimuli (no. 9 and no. 10). The black stripes represent the regions where tactile vibration was applied. The `vibration status' annotation on the left refers purely to the spatial location of the haptic feedback, not to its frequency or intensity. (c) The general interface screen displayed to participants during the experiment.}
\label{exp1}
\end{figure}

Both tests were set based on one group of designed binarized images. Ten raster images with black and white stripes served as the experimental group, with one picture filled with black serving as the control group. The size of all images is 500 by 500 pixels, and the bandwidths of every black or white band in one image are precisely the same. Fig.~\ref{exp1} shows these images as they were displayed on the mobile screen. In Fig.~\ref{exp1_1}, the width of each black and white stripe is equivalent to 1/20 of the image height. Conversely, in Fig.~\ref{exp1_2}, each stripe is substantially wider, with a width that constitutes only 1/10 of the image’s height. Different bandwidths distinguish 10 images, numbered from 1 to 10, and the bandwidths are 1/500, 1/250, 3/500, 1/125, 1/100, 1/50, 3/100, 1/25, 1/20, and 1/10 of the height; the solid black image is numbered No. 0.

\subsection{Design of Experiments}
\subsubsection{Experiment 1. Grating detection Experiment}
To obtain a quantitative measurement, this experiment did not fully adhere to the traditional design of tactile grating acuity tests (GATs). In this experiment, the participants were required to discern the grating width rather than judging the grating orientation. The design of this experiment was informed by the GAT used in the field of tactile perception~\cite{Peli, VCKelly}. 

The participants were asked to make 20 judgments. In each trial, an image was presented in a designated area of the smartphone screen but was obscured by an irrelevant, low-contrast image to prevent visual observation (Fig.~\ref{exp1_3}). The participants were instructed to touch the designated area with their fingers. As shown in Fig.~\ref{exp1}, if the touch occurred within the black-striped area of the grid image, the phone vibrated; if any other area was touched, there was no vibration. The order of image presentation in each experiment followed a pseudorandom sequence, with the 10 grating images appearing once and only once across the 20 trials. For the remaining 10 trials, the participants were shown fully black-filled images.

The participants were tasked with determining whether the presented image contained a grating pattern. During each trial, data were recorded regarding the displayed image, whether the participants made contact, the number of consecutive touches, the total duration of touch, the participants’ judgment, and the three-dimensional (3D) touch trajectory in relation to time and x–y coordinates. The test follows the process of “touch, record data, judge, record data, save data, and switch to the next image” to ensure the reliability of experimental data collection. The experiment was conducted under identical conditions for the S and VI participants.

In the field of visual ability measurement, research on contrast sensitivity has resulted in the development of multiple methods for measuring contrast sensitivity. These include the Cambridge Low Contrast Gratings, Mars Letter Contrast Sensitivity Test, and Pelli-Robson Contrast Sensitivity Chart~\cite{ARDEN, MARS, RPR}.

This experiment was designed with reference to the method of GAT, which is more straightforward than contrast sensitivity measurement and is usually applied to infants or children to detect the presence of parallel lines of decreasing width~\cite{EHR16}.

Using the 11 images designed for this experiment, GAT can be applied to the haptics of humans. The different widths of the lines in the images match the lines on the stimulus surface at different distances in GAT~\cite{EHR17, EHR18}. The convention of the scale of the lines in the images to the grating frequency can be formulated as follows:

\begin{equation}
\label{eq2}
\frac{1}{F} = 2S\cdot k,
\end{equation}

\noindent where $S$ denotes the scale of the lines in the images (1/10, 1/20, ...) and F denotes the image frequency, with cpmm (cpmm = number of cycles per millimeter) as its unit. Notably, the coefficient $k$ is given by

\begin{equation}
\label{eq3}
 k = H_i\cdot \frac{P_H}{P_L},
\end{equation}

\noindent where $H_{i}$ denotes the screen height, $P_{L}$ denotes the screen resolution height, and $P_{H}$ denotes the image resolution width.

In this experiment, 
\begin{align}
k &= 164.9mm \times \frac{720}{1600}. \\
k &= 74.205 mm.
\end{align}
\noindent 

In GAT, the result is reported as `responded to \_\_\_\_ cpmm grating at a distance of \_\_\_\_ cm/inches'. Although standard GATs primarily focus on the behavioral responses of infant subjects to images with varying cpmm values, we consider adult participants. Because adults can articulate their perception of the test stripes and identify their upper sensitivity thresholds, their actual grating acuity can be approximately determined.

Images were displayed pseudorandomly to mitigate potential learning effects.

The above approach aims to derive the finger vibrotactile sensitivity of a specific individual and extract some touch characteristics for two groups of participants.

Table~\ref{table-single} provides details about each image.

\begin{table}[!t]
    \caption{Cycle-per-mm of Images in grating detection Touch Experiment}
    \centering
    \label{table-single}
    \begin{tabular}{cccc}
     \toprule
    Image Number & Bandwidth/pixels & Scale & Cycle-per-mm \\
    \midrule
    No.0 & 0 (solid black) & 0 & / \\
    No.1 & 1 & 1/500 & 3.369 \\
    No.2 & 2 & 1/250 & 2.246 \\
    No.3 & 3 & 3/500 & 1.685 \\
    No.4 & 4 & 1/125 & 1.348 \\
    No.5 & 5 & 1/100 & 0.674 \\
    No.6 & 10 & 1/50 & 0.449 \\
    No.7 & 15 & 3/100 & 0.337 \\
    No.8 & 20 & 1/25 & 0.270 \\
    No.9 & 25 & 1/20 & 0.135 \\
    No.10 & 50 & 1/10 & 0.067 \\
    \bottomrule
    \end{tabular}
\end{table}

\subsubsection{Experiment 2. Grating Different Detection Experiment}
This experiment represents another innovation in this study. The procedure of Experiment 2 is similar to that of Experiment 1, but the images displayed in the designated screen area differ.

\begin{figure}
    \centering
    \subfloat[]{
    \includegraphics[width=0.29\columnwidth]{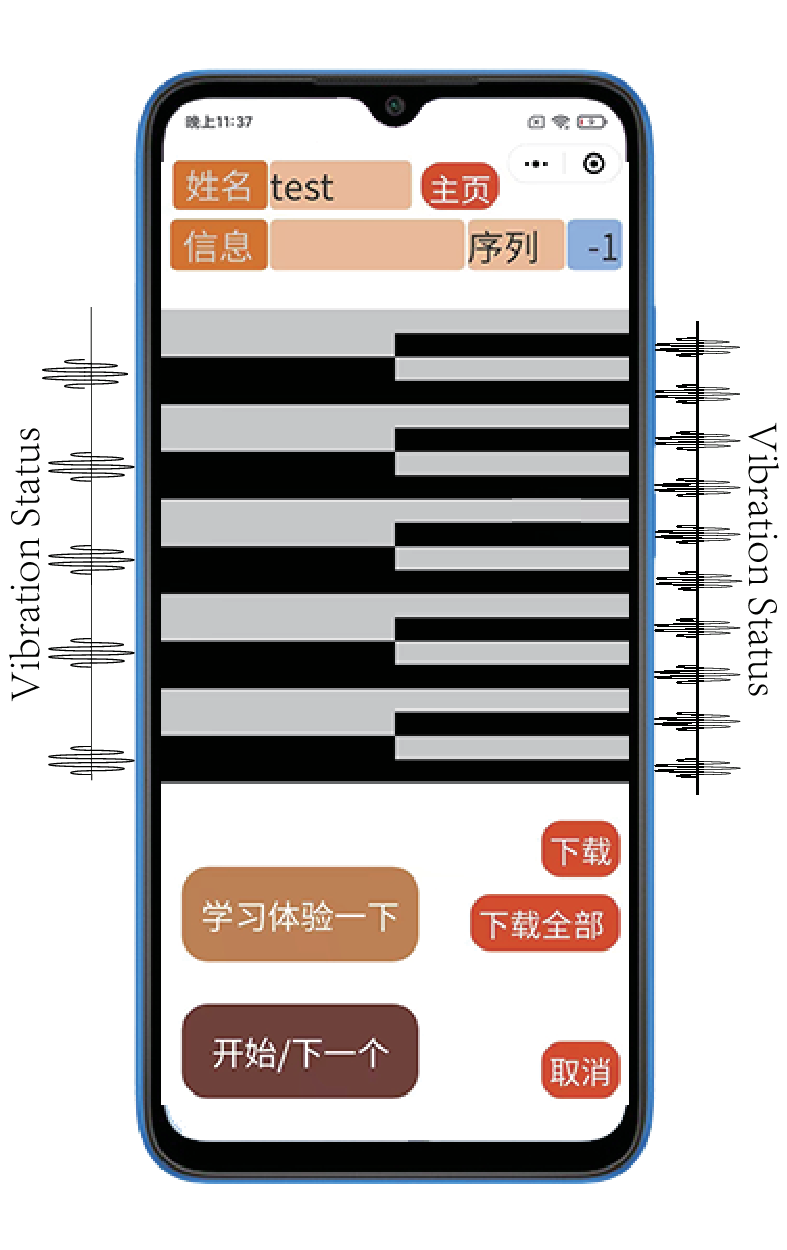}
    \label{exp2_1}}
    \hfil
    \subfloat[]{
    \includegraphics[width=0.29\columnwidth]{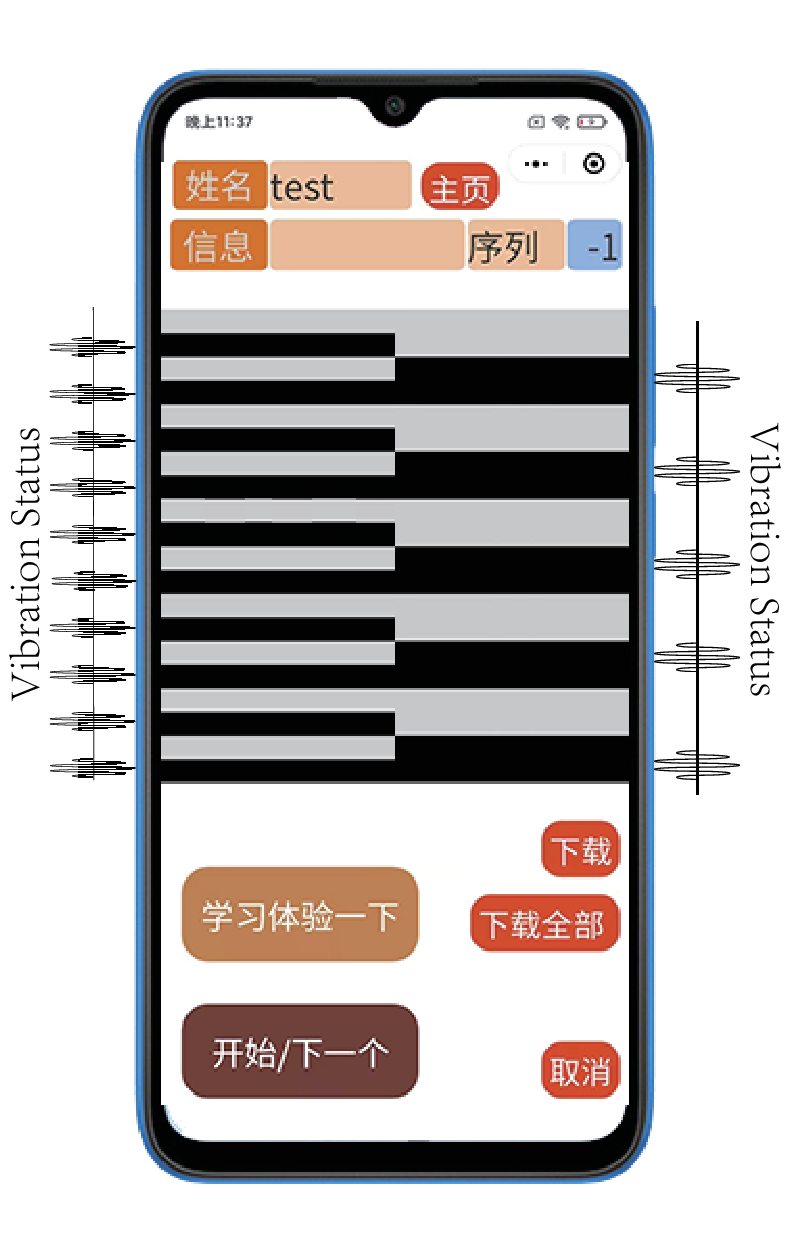}
    \label{exp2_2}}
    \hfil
    \subfloat[]{
    \includegraphics[width=0.31\columnwidth]{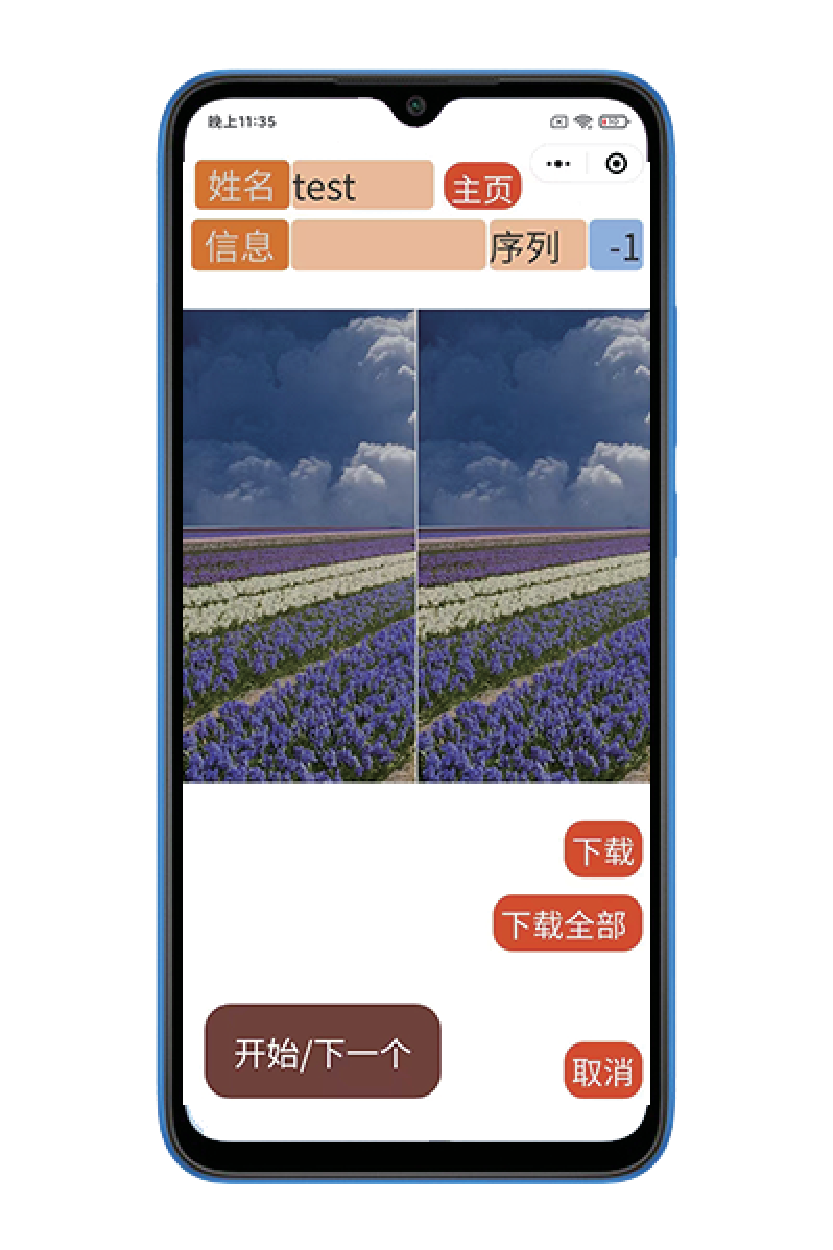}
    \label{exp2_3}}
    \caption{Smartphone interface for Experiment 2. (a) and (b) illustrate two examples of the grating stimuli (no.10+no.9 and no.9+no.10). The black stripes represent the regions where tactile vibration was applied. The `vibration status' annotation on the left refers purely to the spatial location of the haptic feedback, not to its frequency or intensity. (c) The general interface screen displayed to participants during the experiment.}
    \label{exp2}
\end{figure}

In the second set of experiments, the participants were asked to make 20 judgments. In each trial, an image was presented in a designated smartphone screen area but was obscured by an irrelevant, low-contrast image to prevent visual observation (Fig.~\ref{exp2_3}). The participants were instructed to touch the designated area with their fingers. As shown in Figs. ~\ref{exp2_1} and ~\ref{exp2_2}, if the touch occurred within the black-striped area of the grid image, the phone vibrated; if any other area was touched, there was no vibration.

As shown in Fig.~\ref{exp2}, each displayed image comprises a combination of two images, both selected from the 11 images designed for the study. Two grating images of different widths occupy the left and right sides of the displayed image. Only images with neighboring codes were combined, with the fully black-filled image assigned code 0. In addition, combinations presented in reverse order were treated as distinct (Fig.~\ref{exp2}), resulting in 20 unique composite images. Table~\ref{table-double} presents the differences in grating coarseness between the left and right images for each combined image.

The order in which the combined images appear in each experiment follows a pseudorandom sequence. For each trial, data are recorded on the displayed image, including whether the participant made contact, the total interaction duration, and the participant’s judgment. The experiment was conducted under identical conditions for both S and VI participants. The test follows the process of “touch, record data, judge, record data, save data, and switch to the next image” to ensure the reliability of experimental data collection.

Images were displayed pseudorandomly to mitigate potential learning effects~\cite{EHR19, EHR20}. For the sake of reliability of results and presenting the effect of randomness in the outcome, combinations in the reverse order are considered as different combinations and a set of self-comparisons.

The purpose of this experiment is to quantitatively examine the finger vibrotactile discrimination of a specific individual and find a decent comparison interval as a potential option for subsequent haptic feedback mechanisms for VI. This study also aims to explore the characteristics and differences in vibrotactile discrimination between the VI and S individuals. Unlike traditional vibrotactile discrimination experiments, we impose higher demands on participants’ spatial imagination by requiring them to simultaneously touch two objects. When participants touch objects with their eyes closed and then compare them, they engage in spatial processing through tactile spatial imagination. This process involves mentally representing and comparing the shape, size, texture, or position of objects. In this context, participants rely on their ability to construct mental images or representations based on tactile sensory feedback and engage in spatial reasoning to compare the physical properties of objects without visual input~\cite{si2}. The experiment assesses aspects of spatial imagination, including how well individuals can form mental models of objects based solely on tactile input and then mentally manipulate or compare these models.

\begin{table*}[!t]
    \caption{Parameter of Images in Grating Discrimination Experiment}\label{table-double}
    \centering
    \begin{tabular}{ccccc}
     \toprule
    Image Number & Absolute Difference & Percentage Difference & Average Cycle-per-mm & Lg(cpmm)+1.1\\
    & of Bandwidth/pixels & of Bandwidth/\% & (cpmm) & \\
    \midrule
    No.0 \& No.1 or No.1 \& No.0 & 1 & 100 & 6.738 & 1.92853683491616\\
    No.1 \& No.2 or No.2 \& No.1 & 1 & 50 & 2.246 & 1.45141558019649 \\
    No.2 \& No.3 or No.3 \& No.2 & 1 & 33 & 1.348 & 1.22956683058014 \\
    No.3 \& No.4 or No.4 \& No.3 & 1 & 25 & 0.963 & 1.08343879490190 \\
    No.4 \& No.5 or No.5 \& No.4 & 1 & 20 & 0.749 & 0.97429432547683 \\
    No.5 \& No.6 or No.6 \& No.5 & 5 & 50 & 0.449 & 0.75244557586048 \\
    No.6 \& No.7 or No.7 \& No.6 & 5 & 33 & 0.270 & 0.53059682624412 \\
    No.7 \& No.8 or No.8 \& No.7 & 5 & 25 & 0.193 & 0.38446879056588 \\
    No.8 \& No.9 or No.9 \& No.8 & 5 & 20 & 0.150 & 0.27532432114081 \\
    No.9 \& No.10 or No.10 \& No.9 & 25 & 50 & 0.090 & 0.05347557152446 \\
    \bottomrule
    \end{tabular}
\end{table*}

\subsection{Experimental Setup and Participants}
We developed a comprehensive experimental framework to investigate the nuanced aspects of haptic sensitivity in VI individuals compared with their S counterparts. The primary focus of this exploration was on their interaction with vibratory stimuli, which were systematically varied in size to assess a broad range of tactile responses. The vibratory stimuli were presented through a carefully designed interface on a standard smartphone, chosen for its ubiquity and familiarity, ensuring a comfortable and relatable environment for all participants.

To provide a thorough examination of haptic sensitivity, we included two diverse groups of participants: 6 VI individuals (VI group) and 10 S individuals (S group), all aged between 18 and 35 years. This age range was selected because of the pervasive use of smartphones among younger demographics. Further, to minimize potential discrepancies in understanding experimental instructions and general familiarity with research procedures, given that the S group in the baseline testing comprised highly educated researchers, the inclusion criteria for the VI group strictly required a minimum of 12 years of formal education. The inclusion criteria for the VI group were designed to encompass a range of visual impairments, from moderate visual impairment to complete blindness, thereby capturing a wide variety of tactile experiences. This approach enabled us to cover differences in haptic sensitivity across a range of visual capabilities. Our evaluation of the degree of visual impairment follows the IDC11 standard~\cite{idc11}. Further, all participants possessed intact tactile perception

\begin{table*}[!t]
    \caption{Details of the VI Participants}\label{participants}
    \centering
    \begin{tabular}{ccccc}
     \toprule
    Sex & Etiology of Blindness & Residual Vision & Age & Braille Usage \\
    \midrule
    M & Possible Retinoblastoma & None & 21 & Yes\\
    F & Retinoblastoma & None & 19 & Yes \\
    M & Congenital Retinopathy of Unknown Etiology & Light perception & 20 & No \\
    M & Retinopathy of Prematurity & None & 21 & Yes \\
    M & Congenital cataract & Light perception & 22 & Yes \\
    F & Retinitis Pigmentosa & Light perception & 26 & Yes \\
    \bottomrule
    \end{tabular}
\end{table*}

The S group, comprising individuals with no known visual impairments, served as a baseline for typical tactile interactions with vibratory stimuli on a smartphone. This comparative analysis between the VI and S groups was essential for identifying potential differences and similarities in haptic perception, which could inform the design of more inclusive and accessible technology for a diverse user base.

The experimental setup simulated real-world smartphone interactions to ensure the relevance and applicability of the findings. Participants were asked to complete a series of experiments that involved responding to various vibratory stimuli generated by a smartphone to assess their ability to detect and differentiate between stimuli of varying sizes and the speed and accuracy of responses, providing a comprehensive evaluation of haptic sensitivity.

To incentivize conference participation, a food voucher valued at 25 RMB was offered as compensation to each participant.

Some participants were pilot participants and provided feedback on the initial study methods. These participants were excluded from the data analysis. Among the 18 remaining participants, 2 were removed from the analysis due to missing greater than 50\% of their data (S). The remaining participants comprised 4 females (1 VI) and 12 males (5 VI).

The participants were given a brief training period for both grating detection and different detection experiments. The training session was conducted using the same setup and interface as Image No. 6 of Experiment 1. Specifically, the smartphone was secured on a stand, and the physical arrangement was adjusted to ensure that every participant could access the target area using natural, comfortable gestures. During this phase, participants received standardized instructions: (1) to use a single, preferred finger for screen exploration; (2) that haptic feedback is only elicited through active movement over the horizontally oriented stripes located in the screen’s center; and (3) that their core tasks involved either detecting non-vibrating intervals (Experiment 1) or comparing stripe widths (Experiment 2). The training had no fixed duration and concluded when the participant felt sufficiently comfortable with the procedure.

In summary, this study focuses on participants’ acquisition of graphical information through vibratory feedback and their interactive responses to vibratory stimuli with a smartphone interface, using a structured experimental approach. The findings are intended to provide valuable insights into assistive technology, with the aim of enhancing digital accessibility for the VI individuals.

\subsection{Experimental Platform}
In most of the tests, the smartphone “Redmi 9A Dual SIM TD-LTE CN 32GB M2006C3LC,” released in August 2020, was used as the experimental platform~\cite{EHR15}. The software designed to conduct experiments was developed on the Android platform. The phone’s display resolution is 720*160 with the display diagonal as 165.9 mm or 6.5 inches, so the pixel density is 269 PPI. The built-in motor driving the mobile phone’s vibration is “VC0830B009L”~\cite{EHR15}, a coin-type vibration motor that can provide 1.1 Gp Min of vibration strength (testing jig 100g). To ensure ubiquity, some pretests were conducted on more recently produced mobile phones and other operating systems but were not used in formal experiments.

The Android Vibrator API was used, with each distinct vibration event lasting 0.1 s. Concurrently, the getRAWY(index) method was used to record the spatial touch point coordinates, determining whether tactile feedback should be initiated, repeated, or terminated. Further, to ensure consistency across different users, the contact surface’s geometric center was calculated. This algorithm-driven normalization mitigated the influence of fingerpad dimensions.

Based on our experience with the VI community in Shanghai, which comprises approximately 80 individuals, most VI people tend to use entry-level smartphones due to their affordability. Therefore, it is reasonable to select a mobile phone that meets these conditions as the primary experimental platform. Phones such as Redmi 9A, which features a coin-type vibration motor, are a suitable choice. Meanwhile, most latest-generation smartphones are equipped with linear motors. The advantage of the coin-type vibration motor lies in its ability to transmit most vibrations to the front screen surface. As a result, using this type of motor allows for a more accurate reflection of the vibrotactile discrimination when the testers touch the screen with their fingertips.

\section{Results}
\subsection{vibrotactile discrimination Test Results}
Experiment 1 was meticulously designed to investigate the participants’ ability to discern and identify distinct vibratory zones displayed on a smartphone screen. The smartphone interface was programmed to generate vibrations within a specified zone; the zone is divided into bars, characterized by the width of bars, precisely 50, 25, 20, 15, 10, 5, 4, 3, 2, and 1 pixels. These vibratory zones were activated if participants touched the area to challenge their tactile perception skills in a controlled yet realistic simulation of smartphone usage.

\begin{figure}
    \centering
    \subfloat[]{
    
    \includegraphics[width=0.5\columnwidth]{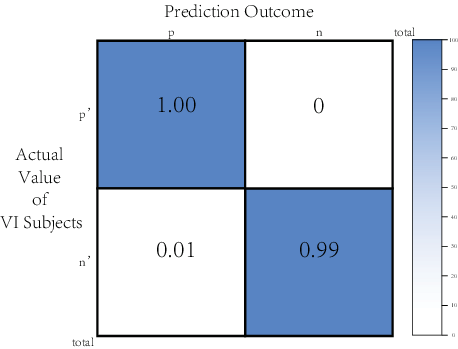}\label{CMB}}
    \subfloat[]{
    \includegraphics[width=0.5\columnwidth]{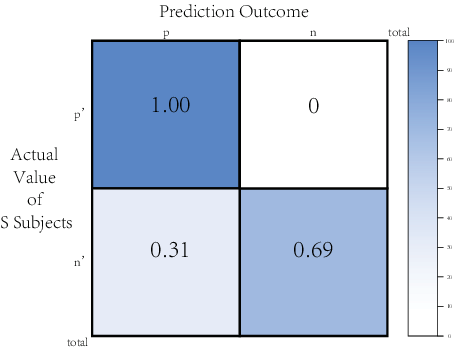}\label{CMS}}
    \caption{Normalized confusion matrices illustrating the tactile classification performance of Visually Impaired (VI) participants (left) and Sighted (S) participants (right). The matrices map the actual stimulus state (rows: p' = positive, n' = negative) to the participants' predicted response (columns: p = positive, n = negative). Values represent the normalized proportion of responses for each condition, with color intensity corresponding to the magnitude as shown in the color bar. VI participants achieved a lower false positive rate (n'/p = 0.01), compared to the S participants (n'/p = 0.31).}
    \label{CMC}
\end{figure}

Fig.~\ref{CMC} shows normalized confusion matrices for two groups of participants, labeled as VI and S participants. Each matrix visualizes the predictive performance of a classification model by comparing actual outcomes (rows) with predicted outcomes (columns). The color intensity corresponds to the normalized values, as indicated by the color bar on the right.

From Fig.~\ref{CMC}, the VI and S participants made accurate judgments for all grating images, meaning they responded to any cpmm grating. According to Table~\ref{table-single} and Eq.~\ref{eq2}, every participant responded to a 3.369-cpmm grating, the highest spatial frequency used in this study. Thus, this experiment did not identify the theoretical upper limit of the tactile grating acuity test for the VI or S participants. However, a comparison of the specific judgment accuracies between the VI and S groups provided indirect evidence of the perceptual potential of touch as a sensory modality and revealed differences in tactile sensitivity between the S and VI participants.

On average, the VI participants demonstrated remarkable proficiency, correctly identifying the vibratory zones with an average accuracy of 99.15\%. In fact, only one participant had a false positive when making a judgment on a picture, as shown on the left side of Fig.~\ref{CMC}. The test achieved a perfect true positive rate (TPR) of 1.00, indicating that all actual positive cases were correctly identified as positive. In addition, the true negative rate (TNR) for negative predictions is remarkably high (0.99), reflecting the VI participants’ ability to accurately classify nearly all true negative cases. Further, the false positive rate is minimal (0.01), and there are no false negatives (FNR = 0), underscoring the VI participants’ overall reliability and precision in evaluating Experiment 1. This high level of accuracy among the VI participants underscores their heightened tactile awareness and sensitivity, possibly a result of their reliance on nonvisual senses for daily navigation and interaction with their environment.

The S participants, although still demonstrating a commendable level of accuracy, scored slightly lower, with an average accuracy rate of 84.5\%. The test achieved a perfect TPR of 1.00, successfully classifying all actual positive cases as positive. However, as shown in Fig.~\ref{CMC}, the TNR is significantly lower (0.69), indicating that 31\% of actual negative cases were misclassified as positive, implying that the S participants adopted a liberal decision criterion and struggled to distinguish between the continuous vibration of a solid black region and the rapid, closely spaced vibratory pulses of high-frequency gratings, resulting in a higher false alarm rate. Although this is not directly related to the vibrotactile discrimination performance of the S participants, it does indicate that the VI participants possess an advantage over those in the S participants in these tasks, as the VI group exhibits significantly higher accuracy in the tests (two-proportion Z-test, $z = 4.24,$ $p$ = $1.1 \times 10^{-5}$).

The S participants performed well on simpler tests, indicating a clear understanding of the basic requirements. However, when faced with tests at a certain difficulty threshold (e.g., fully filled images), they may have encountered limitations in their cognitive strategies. As the S individuals generally do not rely heavily on tactile perception in daily life, they may not have developed an optimal touch-and-judgment strategy quickly. According to signal detection theory, this indicates that the S group employed a relatively lenient criterion— the threshold used to determine whether a signal is present or absent~\cite{SDT}. Signal detection theory also provides a parameter for measuring a participant’s ability to distinguish between signals and noise, known as d-prime~\cite{dp}, which is defined as follows:

\begin{equation}
\label{eq4}
d' = t(H) - t(F),
\end{equation}

\noindent where $H$ denotes the hit rate, and $F$ denotes the false alarm rate. The term $t$ represents a process applied to H and F. In this study, the z-transform was used, and the results of the z-transform can be obtained from the standard normal distribution table.

Using the above method, the d-prime values for the VI and S groups are summarized in Table~\ref{table-t1}.

\begin{table*}[!t]
    \caption{D-prime for S Group and VI Group in Experiment 1}\label{table-t1}
    \centering
    \begin{tabular}{ccccc}
     \toprule
    Group & \multicolumn{2}{c}{VI} & \multicolumn{2}{c}{S} \\
    \midrule
    z-transform & $z(H)=z(1)\approx3.465$ & $z(F)=z(0.017)\approx-2.13$ & $z(H)=z(1)\approx3.465$ & $z(F)=z(0.310)\approx-0.495$ \\
    d-prime & \multicolumn{2}{c}{5.595} & \multicolumn{2}{c}{3.96} \\
    \bottomrule
    \end{tabular}
\end{table*}

The d-prime for the VI group was calculated to be 5.595, indicating a high sensitivity to distinguish signal from noise. The d-prime for the S group was 3.96, suggesting slightly lower sensitivity than the VI group.

These results highlight the differences in signal detection performance between the two groups, with the VI group demonstrating superior sensitivity, as evidenced by a higher $d'$ value. The z-transform effectively quantified these differences by normalizing the hit and false alarm rates.

This experiment is a yes or no detection. The predominance of filled images necessitated a higher confidence level in the participants’ choice, thereby enhancing the reliability of each choice. Although the two groups exhibited similar accuracy rates in the recognition of the grating images, the data, combined with the $d'$ calculation results, indicate that the VI group exhibited greater confidence in their judgments, as they had more accurate identification of all images.

\begin{figure} 
    \centering
    \includegraphics[width=\columnwidth]{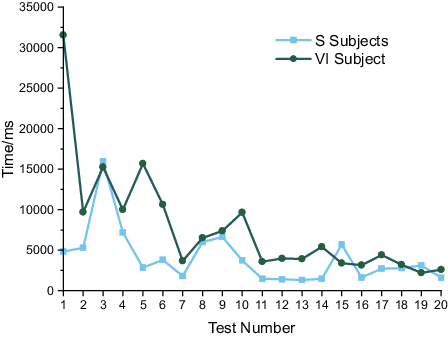}
    \caption{Mean judgment time (ms) per test trial for Visually Impaired (VI) and Sighted (S) participant groups. The graph shows the judgment time for VI participants starting significantly higher but rapidly decreasing, indicating a learning effect and convergence towards the S participant baseline over 20 trials.}
    \label{R311}
\end{figure}

In addition, we considered whether the difference was due to longer and more detailed touching. Fig.~\ref{R311} provides a clear and comprehensive representation of these findings. The comparison visually depicts the detailed results. The difference in judgment time between the two groups decreased as the test progressed. We believe that this is because VI people cannot intuitively understand the test content by looking at the raster pattern schematic, which resulted in the long time taken for the first few judgments. They need to understand what is grating through vibration and touch. The difference in the time required for making judgments between the VI and S groups also aligns with a part of the results of Experiment 2: namely, owing to congenital blindness and the associated limitations in spatial imagination, the VI individuals may not necessarily demonstrate their advantages in vibrotactile discrimination when faced with tests that demand higher levels of spatial reasoning.

\subsection{Combined Sensitivity Analysis}

If the `vibration or no vibration' task in Experiment 1 is considered as a binary judgment, then the tests in Experiment 2 can be viewed as a `grayscale judgment'. The different grating widths provided various `grayscale gradients', posing higher demands on the participants’ overall tactile discrimination abilities.

In an innovative approach for quantifying haptic perception performance, we introduce `contrast sensitivity'  as a new metric through which sensitivity is represented via contrast. This comparative metric aims to provide a comprehensive assessment of participants’ ability to interact haptically, as accurately discriminating between two vibratory stimuli is crucial for understanding haptic interface effectiveness. To achieve this, we combined two raster images with different raster thicknesses, thereby establishing a more distinct sensitivity range for each participant.

According to Table~\ref{table-double}, each pair of images presented in reversed order within an experimental group can be considered as a control group for one another (e.g., No.1 \& No.0 and No.0 \& No.1). Therefore, in data processing, the two responses from a participant to a reversed image pair were treated as a correlated dataset and categorized within the same test for analysis. For the combination of image No. n and No. (n + 1), it is designated as Test n. As shown in Table~\ref{table-double}, the 20 stimuli can be categorized into 10 tests, with the relevant parameters for each test detailed in the table. As a two-alternative forced choice (2AFC) experiment for contrast detection, a participant’s response was considered valid only if they achieved 100\% or 0 accuracy, meaning they provided consistent judgments in both trials~\cite{2ifc,YE}. Conversely, if their accuracy was 50\%, indicating contradictory responses across the two trials—precisely at the chance level—the data were deemed invalid~\cite{50}. In this study, a total of n = 76 data points were excluded, accounting for 23.75\% of the total sample size.

\begin{figure}
    \centering
    \includegraphics[width=\columnwidth]{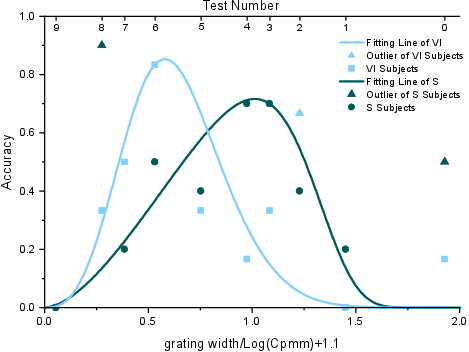}
    \caption{Pyschometric function of distinguishing two gratings with small difference in width. Horizontal axis is offset logarithma of average frequency of the two gratings, log (cycle per mm) +1.1. VI peaked 0.27 cpmm and S subjects peaked at 0.963 cpmm.}
    \label{lou}
\end{figure}

As a 2AFC experiment for contrast detection, the results of this study can be described using a psychometric function, which is a mathematical model that maps stimulus intensity (a parameterized variable) to response accuracy~\cite{pf}. In this study, the average scale of the two images within each test was first calculated, and the corresponding average cycles per millimeter (cpmm) were then derived. Following the mathematical transformation, the stimulus variable was defined as log(cpmm) + 1.1. The dependent variable was the overall judgment accuracy of each group of participants for each test. After sorting the judgment results for each image pair, Fig.~\ref{lou} is obtained. As shown in the figure, we identified two outliers from the original data, one from each group~\cite{out}.These outliers are marked with special symbols. The local outlier factor method is employed to detect outliers in two-dimensional data~\cite{lof}. For a pair of real-number coordinates, this method evaluates the local deviation in density to identify potential outliers. Specifically, for a set of two-dimensional data points, the reachability distance $RD_k(p,o)$ and reachability density $LRD_k(p)$ are calculated for each point.

Then, an outlier factor is derived based on these measures, as follows:

\begin{equation}
\label{eq5}
    LOF_k(p)=\frac{\sum_{o\in N_k(p)}^{}\frac{LRD_k(o)}{LRD_k(p)}}{k},
\end{equation}

\noindent and data points with an outlier factor greater than 1 are classified as outliers.

According to the Fig.~\ref{lou}, the accuracy distribution and outlier patterns of the VI group (n = 6) and the S group (n = 10) can be observed. A modified Weibull function framework was implemented to model the nonmonotonic relationship between stimulus intensity (x) and behavioral accuracy (y) observed in psychophysical measurements. The proposed model integrates dual Weibull components to simultaneously capture ascending sensitivity trends and subsequent suprathreshold decay patterns~\cite{weibull2}:

\begin{equation}
\label{weibull}
y = \mathcal{A} \left(1 - e^{-(x/\beta_1)^{\gamma_1}}\right) \cdot e^{-(x/\beta_2)^{\gamma_2}},
\end{equation}

\noindent where $\mathcal{A}$ denotes the maximum attainable accuracy and governs stimulus detection sensitivity, $\beta_1$ and $\gamma_1$ control performance degradation at high-intensity levels, and $\beta_2$ and $\gamma_2$ control it at low-intensity levels. To ensure biological interpretability and adherence to empirical observations, the model was explicitly constrained to satisfy:

\begin{equation}
\label{weibull_peak}
y(x_{peak}) = y_{max},
\end{equation}

\noindent where $x_{peak}$ denotes the stimulus intensity corresponding to the observed maximum accuracy $y_{max}$ in the experimental data. This constraint was implemented through algebraic reparameterization, eliminating redundant degrees of freedom while preserving model flexibility.

Parameter estimation was conducted under biologically plausible constraints. During the initial parameter design, the value of $\gamma$ was constrained between $0.95x_{peak}$ and $1.3x_{peak}$ to ensure that the decay component would not prematurely suppress the response. The final implementation demonstrated sufficient flexibility to accommodate the specific requirements of the present model fitting. This dual-phase modeling approach enhances conventional Weibull psychometric functions by explicitly accounting for performance decline at high stimulus intensities, a crucial requirement for accurately modeling paradoxical perceptual responses in overdriven sensory systems~\cite{weibull3}.

Analysis indicates that the accuracy distribution of the VI group exhibits right-skewness, with a peak at Test 6 and a maximum accuracy of approximately 0.83. The median accuracy for the VI group was 0.34 (Interquartile Range (IQR) = 0.4167 $-$ 0.0833 = 0.3334). The fitted model yielded a slope parameter of $\beta_1$=0.5000 for the ascending phase and $\beta_2$=0.6650 for the descending phase. Meanwhile, the peak of the accuracy trend for the S group is further left, occurring at Test = 3 and Test = 4, with a maximum accuracy of 0.7. The median accuracy for the S group was 0.4 (IQR = 0.7 $-$ 0.2 = 0.5). The fitted model yielded a slope parameter of $\beta_1$=1.0000 for the ascending phase and $\beta_2$=1.3231 for the descending phase.

From Table~\ref{table-single} and Table~\ref{table-double},  the images in Test 6 correspond to No. 6 and No. 7, with an average cpmm of 0.270 cycles/mm. Further examination of the results reveals that, given the lower median accuracy in the VI group (0.34) than the S group (0.4) and the smaller IQR in the VI group (0.3334) than the S group (0.5), the accuracy distribution in the VI group is more concentrated around lower accuracy values~\cite{IQR}. However, the peak accuracy in the VI group (0.83) is significantly higher than that in the S group (0.7). In addition, the peak range in the VI group is narrower, indicating that its optimal response point (x = 0.53, cpmm = 0.270) demonstrates a higher degree of specificity. This suggests that the frequency associated with the images in Test 6 represents an optimal and easily recognizable tactile stimulus for the VI participants. The nominal base diameter of Braille dots is specified as 1.44 mm (0.057 inches), and the cell spacing shall conform to a nominal center-to-center distance of 2.340 mm (0.092 inches) between adjacent dots within the same cell, measured either horizontally or vertically (but not diagonally), equivalent to 0.264 cycles/mm, which aligns with the findings of this study~\cite{dot}. This consistency implies that the VI participants may have fine-tuned their tactile sensitivity to this frequency through years of practice, resulting in an optimized perception of signals around 0.270 cycles/mm. Meanwhile, the S participants demonstrated only chance-level accuracy (50\%) in Test 6. The peak accuracy for the S participants occurred at Tests 3 and 4, corresponding to images with frequencies of 0.963 and 0.749 cycles/mm, which markedly differed from the results observed in the VI group.

From Table~\ref{table-single} and Table~\ref{table-double}, it is evident that the average frequency of the two images in each test gradually decreases from Test 0 to Test 9, suggesting that the most easily recognizable frequency for S participants is at a higher frequency range than for the VI participants. Further, for all tests from Test 5 and lower (i.e., higher frequency stimuli), the S participants consistently achieved higher accuracy than the VI participants (Fig.~\ref{lou}), suggesting a fundamental difference in tactile processing strategies between the two groups, allowing the S participants to achieve relatively higher accuracy when responding to high-frequency signals.

Overall, in this experiment, the VI participants exhibited a lower median accuracy and a smaller IQR, indicating a stronger central tendency and lower variance, with accuracy concentrated at a lower level. Although their overall accuracy did not surpass that of the S participants, they demonstrated a distinct advantage at a specific frequency (0.270 cycles/mm), where their performance was notably superior to their own accuracy at other frequencies and to the performance of the S participants across all frequencies tested in this experiment.

\subsection{Communication System Analysis}
One of the objectives of this study is to explore the feasibility of a smartphone vibration-based visual aid, making it meaningful to evaluate two experiments from a communication system perspective. In Experiment 1, because participants make a binary decision based on the information contained in a single image, the bit error rate (BER) is used as the evaluation criterion. In Experiment 2, participants make judgments based on the combined information of two images, but it is impossible to know the specific information perceived by each participant from each image individually. Therefore, each judgment is treated as a single piece of information derived from two symbols, and the symbol error rate (SER) is used for evaluation. In addition, using the recorded time data, the average data transfer rate (DTR) of the communication system formed by the two groups of participants and the experimental platform was calculated. The calculation formulas for BER and SER are expressed as follows:

\begin{equation}
\label{eq6}
XER = \frac{N_{error}}{N_{total}},
\end{equation}

\noindent where $X$ means B(bit) or S(symbol), and $N_{error}$ denotes the number of erroneous bits during transmission, with $N_{total}$ representing the total number of bits transmitted. The calculation formulas for DTR are as follows:

\begin{equation}
\label{eq7}
DTR = \frac{Data_{all}}{T},
\end{equation}

\noindent where $Data_{all}$ denotes the total data transferred, and T denotes the time taken.

The results are shown in Table~\ref{cs}.The communication system analysis reveals distinct differences in performance between the VI and S groups across the two experiments, with a focus on SER, BER, and DTR. For the VI group, Experiment 1 lacked error rate data but exhibited a low DTR of 0.00272 bit/s. However, in Experiment 2, the VI group exhibited a high SER of 42.5\% and a BER of 0.83\%, along with an improved DTR of 0.0525 symbol/s. Conversely, the S group demonstrated more stable performance, with an SER of 34.5\% in Experiment 1 and 31.5\% in Experiment 2, while consistently maintaining higher DTRs (0.00493 bit/s in Experiment 1 and 0.03788 symbol/s in Experiment 2). These results suggest that although the S group maintains relatively lower error rates and higher throughput, particularly in Experiment 2, the VI group’s performance is notably impacted by higher error rates, indicating potential areas for system optimization, particularly in error rate reduction to enhance overall data transmission efficiency. This result is consistent with previous findings.

\begin{table*}[!t]
    \caption{Communication System Analysis}\label{cs}
    \centering
    \begin{tabular}{ccccc}
     \toprule
    Group & \multicolumn{2}{c}{VI} & \multicolumn{2}{c}{S} \\
     & Experiment 1. & Experiment 2. & Experiment 1. & Experiment 2. \\
    \midrule
    symbol error rate & / & 42.5\% & / & 31.5\% \\
    bit error rate & 0.83\% & / & 34.5\% & / \\
    Data Transfer Rate & 0.00272bit/s & 0.0525symbol/s & 0.00493bit/s & 0.03788symbol/s\\
    \bottomrule
    \end{tabular}
\end{table*}

\section{Conclusion}

In this study, we investigated the vibrotactile discrimination of VI and S individuals in interpreting smartphone-generated vibratory gratings to explore the feasibility of vibration-based graphical accessibility for the VI. Through two experiments, we demonstrated significant differences in tactile perception between the two groups, aligning with the hypothesis that VI individuals demonstrate the potential to comprehend graphical information through haptic vibration due to their heightened reliance on nonvisual sensory pathways.

In Experiment 1, the VI participants achieved near-perfect accuracy (99.15\%) in detecting grating patterns, significantly outperforming the S group (84.5\%). This underscores the VI group’s refined tactile discrimination capabilities, likely honed through daily dependence on touch. Experiment 2 further revealed distinct processing strategies: the VI participants demonstrated peak performance at 0.270 cycles/mm (83.3\% accuracy), a frequency closely matching Braille dot spacing, whereas the S group peaked at 0.963 cycles/mm (70\% accuracy). Notably, this optimal stimulus frequency of the VI group closely aligns with the common frequency of Braille dots because it suggests that the use of Braille in daily life contributes to the fine-tuned frequency of VI individuals. We speculate that neural plasticity plays a role in this phenomenon.
	
The findings of this study may hold critical implications for developing smartphone-based haptic interfaces. By leveraging vibration frequencies optimized for vibrotactile discrimination—particularly those resonant with Braille-like patterns—a properly designed system could enable efficient access to graphical information for VI users.
	
It is critical to acknowledge that these findings are tied to specific experimental conditions and may not be universally applicable across all tactile experiments or VI individuals.

As a pilot study, we minimized constraints on participant behavior and used standard, fundamental hardware platforms, enhancing the ecological validity of the findings. Although we successfully validated the potential of vibrotactile encoding for graphical communication, limitations such as a small sample size and device-specific vibration motor constraints warrant further exploration. Meanwhile, it is meaningful to consider variables such as the degree and duration of visual impairment and any tactile training. Future research should expand participant diversity, incorporate real-world graphical content, and optimize vibration parameters for broader accessibility.

Prior to this study, researchers observed that VI participants could employ diverse, personalized touch strategies—such as using multiple fingers or pressing the device against their skulls. However, all participants strictly adhered to the standardized protocol (using a single finger with the smartphone secured on a stand) throughout the practice and formal testing phases. Previous studies suggest that multifinger touch and auditory cues can enhance tactile perception. Whether viewed as an intrinsic aspect of vibrotactile discrimination or as natural smartphone interaction habits, personalized behaviors constitute a promising direction for future haptic research.

In conclusion, this study advances the empirical support for vibration-based assistive technologies, offering a scalable and portable solution to enhance digital inclusion for VI populations. By bridging sensory modalities and prioritizing user-centered design, such innovations can redefine accessibility, ultimately improving the quality of life for VI individuals.


\newpage
\vfill

\begin{IEEEbiography}[{\includegraphics[width=1in,height=1.25in,clip,keepaspectratio]{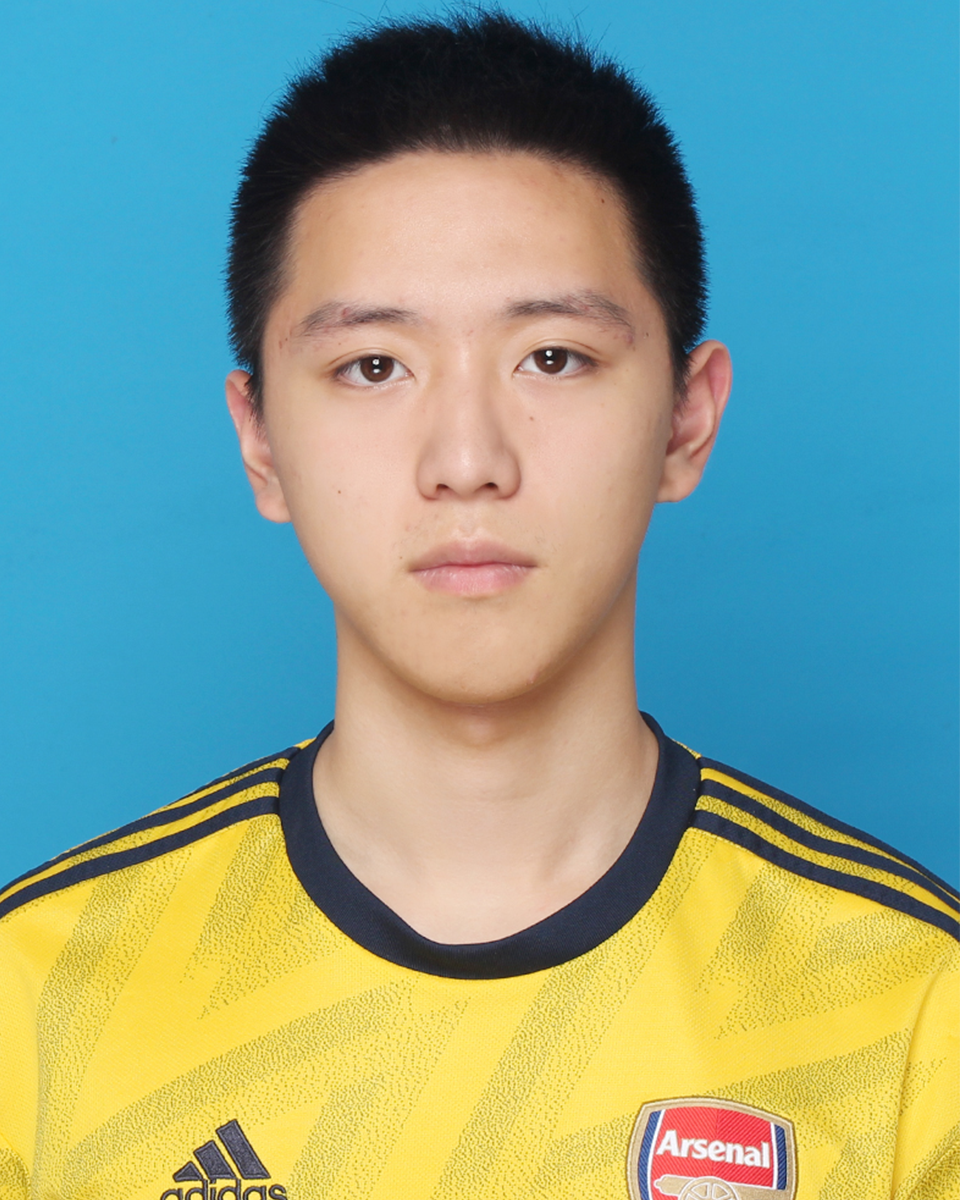}}]{Yichen Gao}
received the B.E. degree from East China Normal University, Shanghai, China. His research interests include interface, perception, and signal processing.
\end{IEEEbiography}
\begin{IEEEbiography}[{\includegraphics[width=1in,height=1.25in,clip,keepaspectratio]{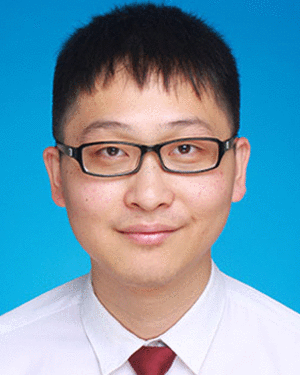}}]{Menghan Hu}
received the Ph.D. degree (Hons.) in biomedical engineering from the University of Shanghai for Science and Technology in 2016. From 2016 to 2018, he was a Post-Doctoral Researcher with Shanghai Jiao Tong University. He is currently an Associate Professor with the Shanghai Key Laboratory of Multidimensional Information Processing, East China Normal University.
\end{IEEEbiography}
\begin{IEEEbiography}[{\includegraphics[width=1in,height=1.25in,clip,keepaspectratio]{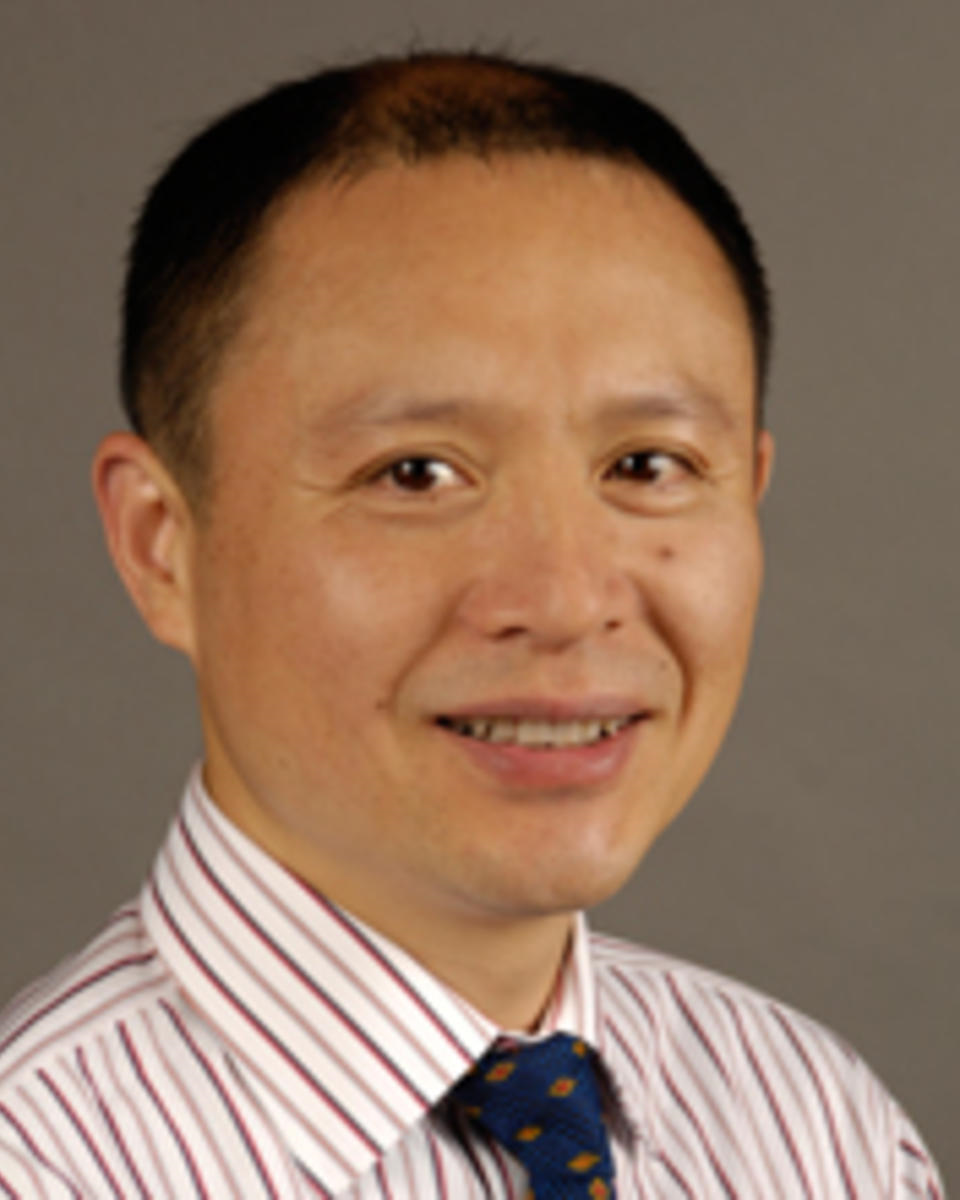}}]{Gang Luo}
received the Ph.D. degree from Chongqing University, China, in 1997. In 2002, he finished his Post-Doctoral Fellow training at the Harvard Medical School, where he is currently an Associate Professor. His primary research interests include basic vision science, image processing, and technology related to driving assessment, driving assistance, low vision, and mobile vision care.
\end{IEEEbiography}

\end{document}